\documentclass[a4paper,fleqn,usenatbib]{mnras}

\usepackage{newtxtext,newtxmath}

\usepackage[T1]{fontenc}
\usepackage{ae,aecompl}

\usepackage{graphicx}	
\usepackage{amsmath}	
\usepackage{amssymb}	
\usepackage{color}

\title[The rapid activity cycle of $\tau$ Boo]{The relation between stellar magnetic field geometry and chromospheric activity cycles II: The rapid 120 day magnetic cycle of $\tau$ Bootis}

\author[S.V.Jeffers et al.]{
S.~V.~Jeffers,$^{1}$
M.~Mengel,$^{2}$
C.~Moutou,$^{3}$
S.~C.~Marsden,$^{2}$
J.~R.~Barnes,$^{4}$  
M.~M.~Jardine,$^{5}$ \newauthor
P.~Petit,$^{6}$ 
J.~H.~M.~M.~Schmitt,$^{7}$
V.~See,$^{8}$
A.~A.~Vidotto,$^{9}$ 
and the BCool collaboration  \thanks{Based on observations made through OPTICON with Telescope Bernard Lyot (TBL, Pic du Midi, France) of the Observatoire Midi-Pyrenees, which is operated by the Institut National des Sciences de l'Univers of the Centre National de la Recherche Scientifique (CNRS) of France.}
\\
$^{1}$ Institut f\"{u}r Astrophysik, Georg-August-Universit\"{a}t, Friedrich-Hund-Platz 1, 37077 G\"{o}ttingen, Germany \\
$^{2}$ University of Southern Queensland, Computational Engineering and Science Research Centre, Toowoomba, 4350, Australia  \\
$^{3}$ CFHT Corporation, 65-1238 Mamalahoa Hwy, Kamuela, Hawaii 9 6743, USA \\
$^{4}$ School of Physical Sciences, The Open University, Walton Hall, Milton Keynes MK7 6AA, UK\\
$^{5}$ SUPA, School of Physics and Astronomy, University of St Andrews, North Haugh, St Andrews, Fife KY 16 9SS, UK  \\
$^{6}$ Institut de Recherche en Astrophysique et Plan\'{e}tologie, 14 Avenue Edouard Belin, F-31400 Toulouse, France  \\
$^{7}$ Hamburger Sternwarte, Universit\"at Hamburg, Gojenbergsweg 112, 21029, Hamburg, Germany \\
$^{8}$ Department of Physics and Astronomy, University of Exeter, Physics Building, Stocker Road, Exeter EX4 4QL, UK \\
$^{9}$ School of Physics,Trinity College, Dublin 2, Ireland 
}

\date{Accepted XXX. Received YYY; in original form ZZZ}

\pubyear{2017}

\begin{document}
\label{firstpage}
\pagerange{\pageref{firstpage}--\pageref{lastpage}}
\maketitle

\begin{abstract}
One of the aims of the BCool programme is to search for cycles in other stars and to understand how similar they are to the Sun. In this paper we aim to monitor the evolution of $\tau$ Boo's large-scale magnetic field using high-cadence observations covering its chromospheric activity maximum.  For the first time, we detect a polarity switch that is in phase with $\tau$ Boo's 120 day chromospheric activity maximum and its inferred X-ray activity cycle maximum.  This means that $\tau$ Boo has a very fast magnetic cycle of only 240 days.  At activity maximum $\tau$ Boo's large-scale field geometry is very similar to the Sun at activity maximum: it is complex and there is a weak dipolar component.  In contrast, we also see the emergence of a strong toroidal component which has not been observed on the Sun, and a potentially overlapping butterfly pattern where the next cycle begins before the previous one has finished.      
\end{abstract}




\section{Introduction} 
The Solar-butterfly diagram illustrates the Solar-cycle both in terms of its 11-year sunspot cycle and its 22 year magnetic cycle.  After the maximum of the cycle, sunspots appear at mid-latitudes, and then evolve closer and closer to the equator until solar minimum is reached. The wings of the Solar butterfly are well separated meaning that the new cycle starts only when the old cycle has finished.  The magnetic cycle of the Sun is aligned with its chromospheric activity (S-index) cycle with polarity reversals of its large-scale field every 11 years.  Currently, the only other star known to show this Solar-like behaviour is the K5 dwarf 61 Cyg A where the polarity cycle of its large-scale field is in phase with its 7.3 year S-index activity cycle \citep{BoroSaikia2016}.  Similar to the Sun, 61 Cyg A's large-scale field is complex at activity maximum and simple at activity minimum.  In contrast, the K2 dwarf $\epsilon$ Eridani shows a complex and highly variable large-scale magnetic field at its activity minimum \citep[][paper I of this series]{Jeffers2017}.

Another star that shows frequent polarity switches is the planet-hosting F7 dwarf $\tau$ Bootis \citep{Catala2007,Donati2008,Fares2009,Fares2013,Mengel2016}.  Surprisingly, $\tau$ Boo's S-index cycle is only 120 days \citep{Mengel2016,Mittag2017} which is three times faster than the yearly polarity reversal frequency derived by \cite{Fares2009}, though they did find weaker evidence for a shorter cycle \citep[and is also discussed by][]{See2016}.  This led \cite{Mengel2016} to suggest that the discrepancy between the two cycle lengths is because the observational epochs used to reconstruct $\tau$ Boo's large-scale magnetic field have been taken too far apart.   If $\tau$ Boo's large-scale field changes polarity every 120 days, or 240 days to return to the original polarity, this would make it the fastest magnetic cycle ever observed. In addition, this would imply that $\tau$ Boo has, similar to the sun or 61 cyg A, polarity reversals that are in phase with its S-index cycle. If confirmed, this would open up a new parameter space for understanding the magnetic field generation processes in the Sun and other Solar-type stars.
The aim of this paper is to determine with certainty if $\tau$ Boo has a shorter magnetic cycle period. If this is the case, it would make $\tau$ Boo an ideal target to compare the evolution of its large-scale field at activity maximum to the Sun's field at activity maximum.  To achieve this we densely monitor the evolution of $\tau$ Boo's large-scale magnetic field geometry and polarity over the S-index maximum that occurred in June 2016.  We reconstruct the geometry and polarity  of $\tau$ Boo's large-scale field: (1) 2 months before S-index maximum to get the polarity of the field before activity maximum, (2) during and immediately after the S-index maximum,(3)1-2 weeks after S-index maximum and (4) 3-4 weeks after S-index maximum.

\begin{table}
\caption{Journal of Observations. The rotational phase is defined as HJD 2453450.984 + 3.31245E and is used for all epochs, with subsequent epochs taking phase = 0 as an integer number of cycles from this value. The exposure time of all observations is 4*600s for the NARVAL Observations (N in column 2)  and 4*90s for the ESPADONS observations (E in column 2). The Julian date shown is +2452000.  The full table is online only.}
\protect\label{t-obslist1}
\begin{tabular}{l c c c c c c}
\hline
\hline
Date  & Ins& Julian  & UT & Phase & LSD & S-\\
2016 & &  Date &&& S/N & index\\
\hline
\multicolumn{7}{c}{Map 1 (2016.21)} \\
17.03 & N    &  464.538  &   00:48:35 & 0.65 &  34130  & 0.187 \\
17.03 & N    &  464.659  &   03:42:58 & 0.69 &  35871  & 0.187 \\
18.03 & N    &  465.570  &   01:35:23 & 0.96 &  33827  & 0.194 \\
.....\\
\hline
\hline
\end{tabular}
\end{table}

\begin{table}
\caption{Stellar Parameters}
\protect\label{t-stparam}
\begin{tabular}{l c c }
\hline
\hline
Parameter & Value & Reference\\
\hline
Effective Temperature (K) & 6399 $\pm$ 45 & \cite{Borsa2015} \\
Mass (M$_\odot$) & 1.39 $\pm$ 0.25 & \cite{Borsa2015} \\ 
Radius (R$_\odot$) & 1.42 $\pm$ 0.08 &  \cite{Borsa2015} \\
$v$ sin $i$ (km s$^{-1}$) & 14.27 $\pm$ 0.06 & \cite{Borsa2015} \\
P$_\mathrm{rot}$ (day) & 3.1 $\pm$ 0.1 & \cite{Mengel2016} \\
Inclination (deg) & 44.5 $\pm$ 1.5 & \cite{Brogi2012} \\
Age (Gyr)  & 0.9$ \pm$ 0.5 & \cite{Borsa2015}\\
\hline
\hline
\end{tabular}
\end{table}

\section{Observations and Data Analysis}

The observations of $\tau$ Boo were secured using the high-resolution echelle spectrographs ESPADONS, located at the Canada France Hawaii telescope, and  NARVAL, located at Telescope Bernard Lyot, France \citep{Auriere2003} in spectropolarimetric mode.  The instrumental setup, observing and data reduction procedure has previously been described in detail by \cite{Mengel2016}.   We observed $\tau$ Boo over a time span of 4 months from March 2016 to July 2016.  The observations are summarised in Tab.~\ref{t-obslist1}. 
All Stokes I and Stokes V reduced spectra were processed using LSD \citep[Least-Squares Deconvolution][]{Donati1997}.  The line mask used was downloaded from the VALD atomic database for a star with: $T_{\rm eff}=6250$K, $\log g=4.0$, a depth threshold of 0.4 and containing an average of 3840 (ESPADONS) and 3716 lines (NARVAL). We use this mask for consistency with \cite{Mengel2016}.  This line mask was applied to each spectrum with a step size of 1.8 km s$^{-1}$ in the LSD profile.  More details on the data processing of $\tau$ Boo are presented in \cite{Mengel2016}.
The S-index values calculated for all spectra are calibrated to the values from the Mount Wilson S-index survey. The S-index values show the highest values over a time period of approximately seven days centered on 9 June 2016.  The S-index values obtained in this work are consistent with the values reported by \cite{Baliunas1997,Mengel2016,Mittag2017,Schmitt2017}.  The first observations in this work occur immediately after the phase jump reported by \cite{Mittag2017}, though we do not notice any effect of this in our S-index measurements.

\section{Large-scale magnetic field geometry}

The large-scale magnetic field geometry of $\tau$ Boo is reconstructed using the tomographic technique of Zeeman-Doppler imaging (ZDI).  The technique samples the stellar surface using a series of local Stokes V profiles which are computed based on the stellar parameters shown in Tab. ~\ref{t-stparam}.  These profiles are then used to compute disk integrated synthetic Stokes V profiles, which are iteratively fitted to the observed Stokes V profiles using the maximum entropy algorithm described by \cite{skilling1984}.  The large-scale magnetic field of $\tau$ Boo is reconstructed by assuming that the field geometry is projected onto a spherical harmonics frame \citep{Donati2006}, where the magnetic energy is decomposed into poloidal and toroidal components.  A spherical harmonics expansion with a maximum value of $\ell_{max} = 9$ was used as there was no improvement to the fits using larger values.  A reduced $\chi ^2$ of 0.9 was obtained for all of the maps when differential rotation was included in the image reconstruction process.  We determine the differential rotation of the magnetic features using the same method as previously applied to $\tau$ Boo by \cite{Mengel2016}.  The resulting values are tabulated for each epoch in Tab.~\ref{t-diffrot}, where the values for Map 2 are also used in the reconstruction of Map 1 due to its sparser phase coverage.  The values of differential rotation are consistent with the previous measurements of \cite{Donati2008, Fares2009, Mengel2016}.  For all epochs, the equatorial rotation period, $\Omega_{eq}$, remains constant within the errors, and is consistent with the photometric rotation period taking $\tau$ Boo's inclination into account.  In contrast, the difference between pole and the equator, d$\Omega$, is shown to vary from 0.10 $\pm$ 0.04 rad d$^{-1}$ in May 2014 to 0.5 $\pm$ 0.12 rad d$^{-1}$ in June 2007. 

\begin{figure*} 
\includegraphics[width=0.99\textwidth]{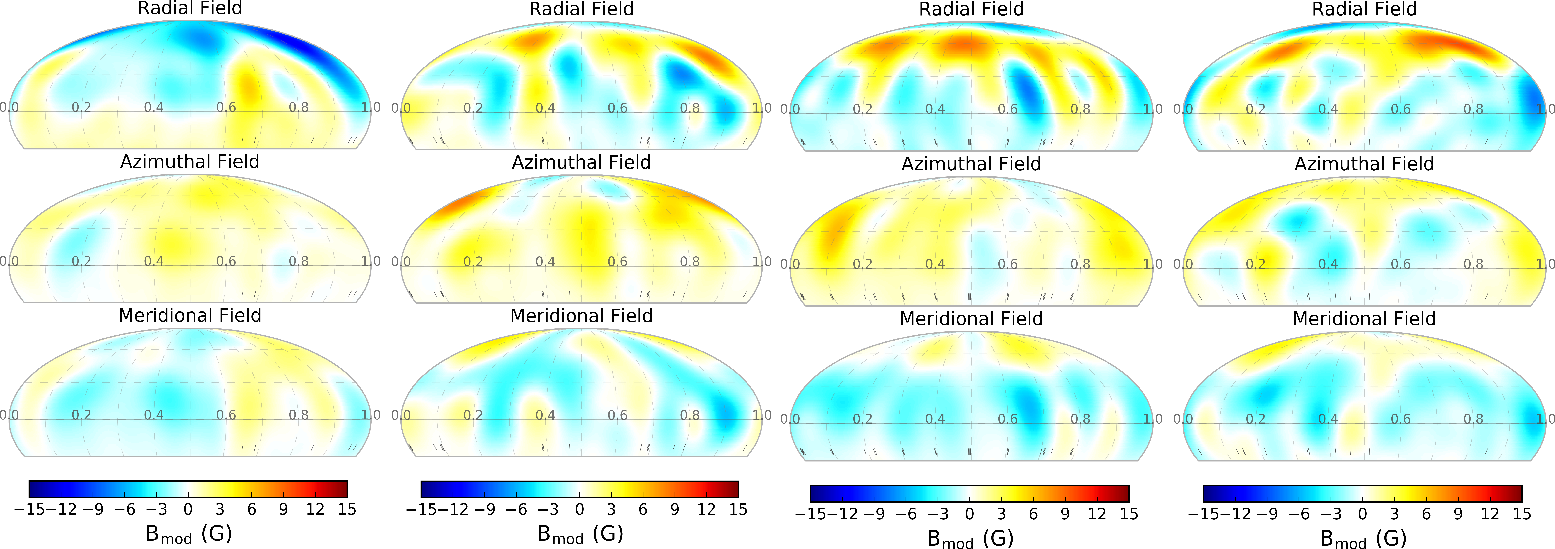}
\vspace{-0.2cm}
\caption{Magnetic field maps of $\tau$ Boo reconstructed for (left to right) Map 1 (2016.21), Map 2 (2016.44), Map 3 (2016.47) and Map 4 (2016.54).  Each image has the same scale with B$_{max}$=15G and where yellow/red indicates positive polarity and cyan/blue negative polarity. The tick marks at latitude -25$^\circ$ indicate the observational phases used in the image reconstruction.} 
\protect\label{f-magmaps} 
\vspace{-0.2cm}
\end{figure*}

\begin{figure*} 
\begin{center}
\includegraphics[width=0.8\textwidth]{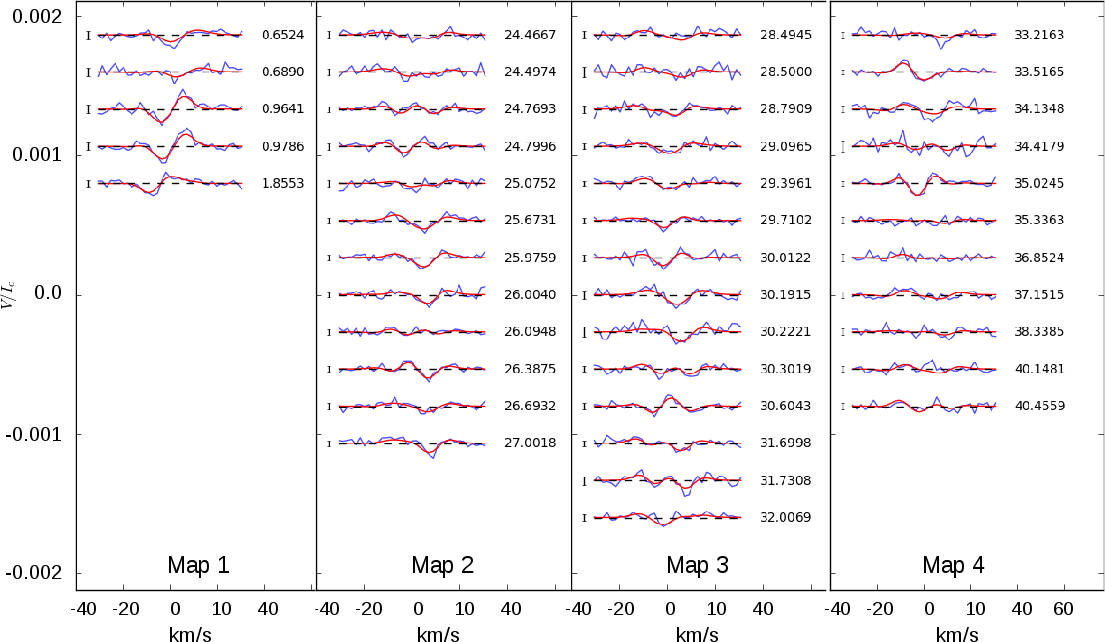}
\caption{Fits to the Stokes V profiles for for each ZDI Maps showing the fits (red line) to the data (blue line).} 
\protect\label{f-magfits} 
\end{center} 
\vspace{-0.7cm}
\end{figure*}

\subsection{Magnetic maps}

The large-scale magnetic field geometry of $\tau$ Boo is shown in Fig. ~\ref{f-magmaps} for four epochs spanning a period of four months.  The fits to the Stokes V profiles are shown in Figure ~\ref{f-magfits}.  The observations were sub-divided up as indicated in Tab.~\ref{t-obslist1} to ensure that any gaps in the  observations were minimised and to ensure that the phase coverage doesn't exceed several stellar rotations due to the strong surface shear.  The division of the observations into the magnetic maps was extensively tested with different combinations of observations.  For example, when we reconstructed five magnetic maps instead of four, the phase coverage was too sparse to reconstruct reliable maps.  The phase coverage of Map 1 is not optimal due to unfortunate weather conditions and we caution a detailed interpretation of this result.  However, the global structures, e.g. the polar spot with negative polarity, are considered to be reliable.

\begin{table}
\caption{Differential Rotation}
\protect\label{t-diffrot}
\begin{tabular}{l c c}
\hline
\hline
Epoch & $\Omega_{eq}$  (rad d$^{-1}$) & d$\Omega$ (rad d$^{-1}$) \\
\hline
Map 1 / 2016.21 & \multicolumn{2}{c}{used Map 2 values} \\
Map 2 / 2016.44 &  2.00 $\pm$ 0.03 & 0.31 $\pm$ 0.08 \\
Map 3 / 2016.47 &  1.99 $\pm$ 0.02 & 0.11 $\pm$ 0.08 \\
Map 4 / 2016.54 &  1.95 $\pm$ 0.01 & 0.16 $\pm$ 0.04 \\
\hline
\end{tabular}
\end{table}

\subsubsection{Radial field component}

Over the time span of the observations there is significant evolution of $\tau$ Boo's radial magnetic field. For Map 1 (2016.21), there is a large polar region of negative polarity. The next epoch is $\sim$2.5 months later at 2016.44 (Map 2) which is near S-index maximum, where we observe a polarity switch and the dominant radial magnetic field at high latitudes now has a positive polarity, with a small weak negative spot at the pole. The new positive magnetic features form a band at high latitudes with stronger spots at phases 0.3 and 0.95.  Additionally, there are low latitude spots with negative polarity at phases 0.25, 0.45 and 0.8.  One week later, Map 3 shows that the previously observed band of positive polarity is still present but more concentrated into spots at 0.15, 0.45 and a weaker structure at 0.75 which had negative polarity one week earlier. These spots also occur at slightly lower latitudes than Map 2 along with a stronger and larger polar spot with negative polarity.  In the final map, one week later, Map 4, the negative polar region has grown in strength and now extends to equatorial latitudes at phase 0.0.  The band of positive polarity at high latitudes now occurs at slightly lower latitudes and the spots are concentrated at phases 0.2 and 0.7--0.95.  There are some weak negative magnetic spots at equatorial phases and a stronger negative spot at phase 1.0.

\subsubsection{Azimuthal field component}

The azimuthal field component remains weak compared to the radial field at all epochs.  At the start of the observations, in Map 1, there is virtually no azimuthal field.   Near to activity maximum a weak positive field starts to emerge.  One week later, there are additionally a few very weak negative spots.  Two weeks after activity maximum the azimuthal field remains weak with several weaker spots of negative polarity.  These results are in contrast to the previous results of ~\cite{Fares2009} where they report a polarity switch in all 3 components, though they note that the azimuthal field lags the radial field by a factor of 18\% (= 22 days assuming a 240 day magnetic cycle).  Since there are negative regions emerging it could indicate that the azimuthal field leads, and the radial field follows.  Additional observations over a full cycle are required to confirm this.

\subsubsection{Meridional field component}

The meridional field remains weak over the time-span of the observations.  As previously discussed by ~\cite{Donati1997a}, the cross talk between the radial and meridional fields primarily effects magnetic features at low latitudes, which implies that the higher latitude features are reliably reconstructed. Given that $\tau$ Boo's inclination is 44.5 $\pm$ 1.5$^\circ$ \citep{Brogi2012} the direction of the cross-talk, e.g. meridional to radial or from radial to meridional, is not clear. 

\subsection{Magnetic energy}

\begin{figure} 
\begin{center}
\includegraphics[scale=0.31]{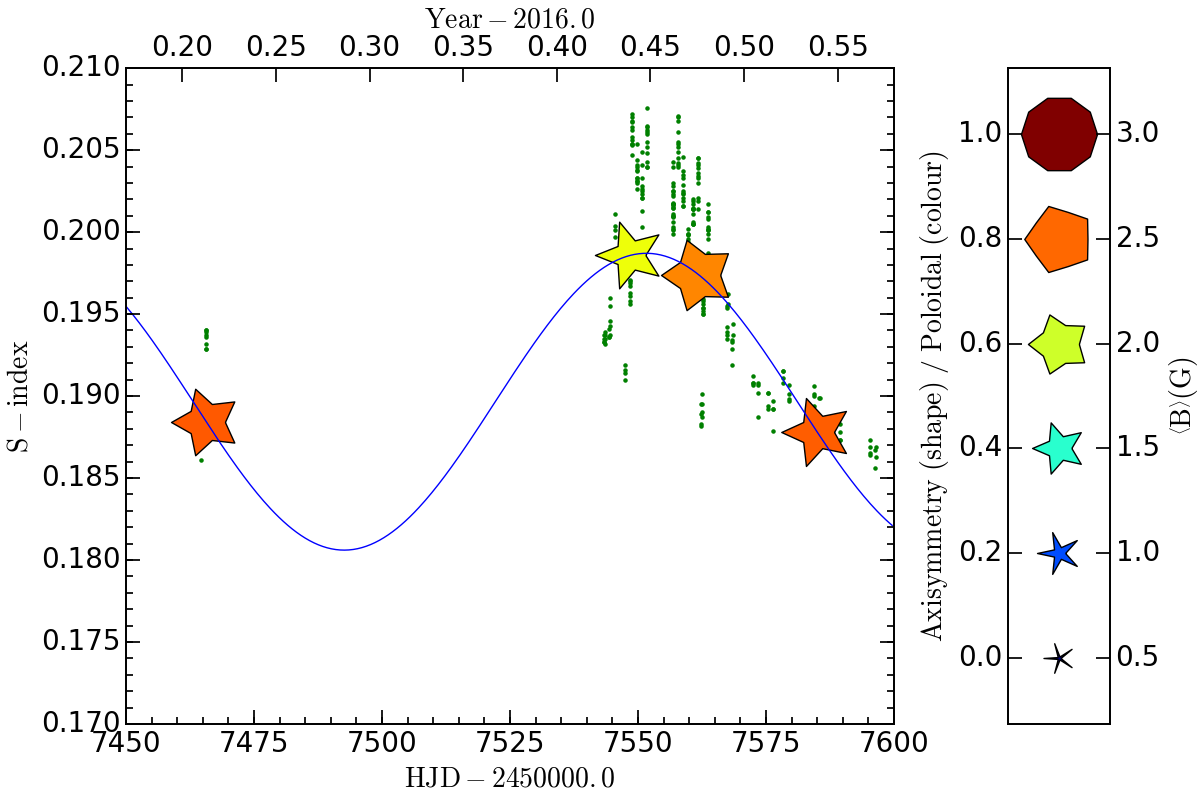}
\caption{The evolution of $\tau$ Boo's large-scale field during S-index maximum.  The symbol shape indicates the axisymmetry of the field (non axisymmetric by pointed star shape and axisymmetric by decagon), the colour of the symbol indicates the proportion of poloidal (red) and toroidal (blue) components of the field and the symbol size indicates the magnetic field strength.  The green points are the individual S-index measurements.  The blue line indicates the S-index cycle and is a continuation of the S-index cycle from \protect\cite{Mengel2016}.} 
\protect\label{f-confusogram} 
\end{center} 
\end{figure}

Over the time-period of the observations, $\tau$ Boo's large-scale magnetic field exhibits a polarity switch before activity maximum, and a notable evolution in its large-scale field after activity maximum (Maps 2, 3 and 4).  In terms of the magnetic energy, the components are summarised in Tab.~\ref{t-mag_en} and in Fig.~\ref{f-confusogram}.  Over the timespan of all four maps, there is an increase in the field complexity for orders $\ell>$ 3, from 23\% in Map 1, though this could result from the sub-optimal phase coverage, to 55\% in Map 4.  At the same time, there is a notable re-emergence of the toroidal field energy after the polarity switch in Map 2 (36\%) which then decreases back to the level of Map 1 by Map 4.  There is also a sharp decrease in the dipolar component of the poloidal field at activity maximum which then increases.  Finally, there is a likely increase in the toroidal component of the axisymmetric field after the polarity switch, from 57\% in Map 2 to 63\% in Map 3, and then rapidly decreases to 31\% in Map 4.  The other noteworthy change is the fraction of energy in the axisymmetric component where there are small fluctuations for all maps.  The remaining components of the magnetic field energy show small fluctuations, though the average field strength remains constant over all of the maps.

\begin{table*}
\caption{Fraction of the large-scale magnetic energy reconstructed in the toroidal and poloidal field components; the fraction of the poloidal field in the dipolar ($\ell$=1), quadrupolar ($\ell$=2) and octupolar ($\ell$=3) components; and the fraction of the energy stored in the axisymmetric component ($m$=0). Caution is advised in a detailed interpretation of the results for Map 1 due to its poor phase coverage.}
\vspace{-0.8cm}
\protect\label{t-mag_en}
\begin{center}
\begin{tabular}{l l c c c c c c c c c c c c }
\hline
\hline
Map / Epoch & Bmean & Bmax & Toroidal & Poloidal & Dipolar & Quad. & Octupolar & $\ell>$3 & Axisym. & Poloidal & Toroidal \\ 
 & (G) & (G) & (\% tot) & (\% tot) & (\% pol) & (\% pol) & (\% pol) & (\% pol) & (\% tot) & (\% axi) & (\% axi) \\
\hline
1 / 2016.21 & 2.4 & 12.0 & 15 & 85 & 50 & 15 & 12 & 23 & 51 & 55 & 28 \\
2 / 2016.44 & 2.5 & 9.2 & 36 & 64 & 9 & 12 & 16 & 37 & 38 & 27 & 57  \\
3 / 2016.47 & 2.8 & 9.4 & 23 & 77 & 28 & 9 & 14 & 51 & 56 & 53 & 63  \\
4 / 2016.54 & 2.6 & 9.9 & 19 & 81 & 23 & 11 & 11 & 55 & 42 & 44 & 31 \\
\hline
\hline
\vspace{-0.8cm}
\end{tabular}
\end{center}
\end{table*}

\begin{figure*} 
\begin{center}
\includegraphics[width=0.7\textwidth]{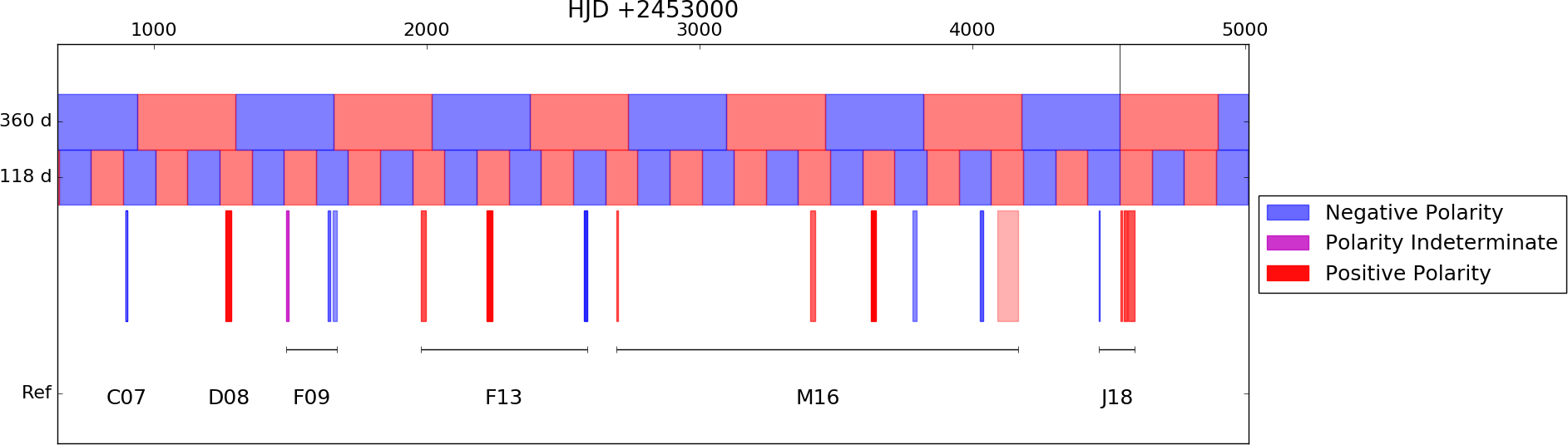}
\caption{Schematic illustration of the correlation of the polarity of the reconstructed ZDI maps (bottom row) with the 1 year (top row) and 118 day (middle row) magnetic cycle periods.  Note that all the observed polarities match the schematic polarity only for the 120day cycle.  References: \protect\cite{Catala2007,Donati2008,Fares2009,Fares2013,Mengel2016}.  }
\protect\label{f-polswitch} 
\end{center} 
\vspace{-0.7cm}
\end{figure*}

\section{Discussion}

Our results, together with previous observations, show with certainty that the polarity of $\tau$ Boo's large-scale magnetic field changes on a time-scale of 120 days and is co-incident with its S-index cycle.  The frequency of the polarity switch is compatible with the polarity of previous ZDI maps as shown in Figure ~\ref{f-polswitch}.  $\tau$ Boo is only the second star, apart from the Sun, known to have a polarity switch that is coincident with its S-index cycle.  The 120-day S-index cycle of $\tau$ Boo is much faster than the 7.3-yr cycle of 61 Cyg A and the 11-yr cycle of the Sun.

Over the short time-span of our observations, $\tau$ Boo's large-scale magnetic field shows large variations including a negative to positive polarity switch that happened before its S-index maximum.  The radial field evolves significantly over a timescale of a few weeks following the polarity switch, where there is a surprising emergence of a polar spot of negative polarity.  In contrast, the azimuthal field shows a complex, though weak, field structure which is dominated by regions of positive polarity, though regions of negative polarity start to emerge by Map 4.   These results are comparable to epochs May 2014 and January 2015, which were also observed after activity maximum \citep{Mengel2016}.  

Our results show that there are important characteristics of $\tau$ Boo's large-scale magnetic field that are both comparable and, at the same time, different to the large-scale magnetic field of the Sun.  The first similarity is the large increase in the field complexity with more energy in modes $\ell>$ 3 which ranges from 37\% at activity maximum (Map 2) and then up to 55\% over the time-span of 1.5 months.  An increase in the complexity of the large-scale field at activity maximum has previously been observed on the Sun and the K dwarf 61 Cyg A \citep{DeRosa2012,BoroSaikia2016}.  The second similarity is that there is a strong decrease in the dipolar component after the polarity switch, which then rises again over the following 1.5 months after activity maximum.  The Sun shows a similar trend as reported in Fig. 1 of \cite{DeRosa2012}, though over a longer time period than that for $\tau$ Boo.  Interestingly, after the polarity switch, there is an emergence of a significant toroidal field which decreases during the 1.5 months after activity maximum.  Currently there are no comparable observations of the Sun's toroidal field but if the Sun's large-scale field is similar to $\tau$ Boo's large-scale field, it should exhibit similar field characteristics.  An analysis using vector synoptic maps of the Sun would be very insightful \citep{Vidotto2016}.  In the case of the Sun, the quadrupolar field is an order of magnitude larger than the dipolar field, which is in contrast to what we observe on $\tau$ Boo.  The main polarity switch that we have observed is from negative to positive polarity field that occurs just before activity maximum.  The emergence of an additional magnetic spot with negative polarity after activity maximum (e.g.at polar regions in Maps 2 to 4) resembles an overlapping butterfly diagram, where one cycle starts before the previous one has finished, in contrast to the well separated butterfly diagram observed on the Sun.  More observations of $\tau$ Boo are needed to confirm this.

Another characteristic feature of the Solar-cycle is that the X-ray magnetic cycle is aligned with its S-index cycle. Previous X-ray observations of $\tau$ Boo by \cite{Poppenhaeger2012}  did not detect a coronal magnetic cycle.  The reason why a cycle was not detected is likely because the X-ray measurements were secured at very sparsely sampled epochs. An alternative explanation could be that magnetic cycles are not always observable in X-rays \citep{McIvor2006}. Additional X-ray detections of $\tau$ Boo, recently reported by \cite{Mittag2017}, could be in phase with $\tau$ Boo's S-index cycle, though further higher cadence X-ray observations are required to confirm this indicative coronal cycle.  Our results show that a polarity switch in $\tau$ Boo's large-scale magnetic geometry field coincides with its 120 day S-index and inferred coronal activity maximum and opens up a new parameter space in the quest for understanding the magnetic field generation mechanisms in the Sun and other Solar-like stars.

\section*{Acknowledgments}

We would like to thank Sudeshna Boro Saikia and Rim Fares for valuable contributions in discussing the results of this paper, and their contributions to the observing proposal. SJ acknowledges support from the German Science Foundation (DFG) Research Unit FOR2544 ``Blue Planets around Red Stars'', project JE 701/3-1.

\bibliographystyle{mnras}
\bibliography{iau_journals,tauboo}

\bsp	
\newpage
\appendix

\begin{table}
\caption{Journal of Observations.  The rotational phase is defined as HJD 2453450.984 + 3.31245E and is used for all epochs, with subsequent epochs taking phase = 0 as an integer number of cycles from this value. The exposure time of all observations is 4*600s for the NARVAL Observations (N in column 2)  and 4*360s for the ESPADONS observations (E in column 2). The Julian date shown is +2452000.}
\protect\label{t-obslist}
\begin{tabular}{l c c c c c c}
\hline
\hline
Date  & Ins& Julian  & UT & Phase & LSD & S\\
2016 & &  Date &&& S/N & index\\
\hline
\multicolumn{7}{c}{Map 1 (2016.21)} \\
17.03 & N    &  464.538  &   00:48:35 & 0.65 &  34130  & 0.187 \\
17.03 & N    &  464.659  &   03:42:58 & 0.69 &  35871  & 0.187 \\
18.03 & N    &  465.570  &   01:35:23 & 0.96 &  33827  & 0.194 \\
18.03 & N    &  465.618  &   02:44:17 & 0.98 &  36018  & 0.193 \\
21.03 & N    &  468.522  &   00:25:47 & 1.86 &  46888  & 0.189 \\
\\
\multicolumn{7}{c}{Map 2 (2016.44)} \\
03.06 & N    &  543.422  &   22:02:33 & 24.47 &  36728  &	0.194 \\
04.06 & N    &  543.524  &   00:28:55 & 24.50 &  39730  &	0.194 \\
04.06 & N    &  544.424  &   22:05:54 & 24.77 &  46105  &	0.194 \\
05.06 & N    &  544.525  &   00:30:31 & 24.80 &  43656  &	0.195 \\
05.06 & N   &  545.437  &   22:24:50  & 25.08 &  40260  &	0.200 \\
07.06 & N    &  547.418  &   21:57:11 & 25.68 &  48636  &	0.192 \\
08.06 & N    &  548.421  &   22:01:27 & 25.98 &  52629  &	0.196 \\
08.06 & N    &  548.514  &   24:15:47 & 26.00 &  51456  &	0.197 \\
09.06 & E & 548.81 & 07:28:33 & 26.10 & 50810 & 0.201 \\
10.06 & E & 549.78 & 06:45:02 & 26.39 & 50991 & 0.199 \\
11.06 & E & 550.80 & 07:03:26 & 26.69 & 46764 & 0.197 \\
12.06 & E & 551.82 & 07:35:31 & 27.00 & 49895 & 0.201 \\
\\
\multicolumn{7}{c}{Map 3 (2016.47) } \\
17.06 & E & 556.76 & 06:16:00 & 28.49 & 36576 & 0.197 \\
17.06 & E & 556.78 & 06:42:15 & 28.50 & 29335 & 0.197 \\
18.06 & E & 557.75 & 05:49:39 & 28.79 & 38711 & 0.201 \\
19.06 & E & 558.76 & 06:07:30 & 29.10 & 51118 & 0.198 \\
20.06 & E & 559.75 & 05:56:51 & 29.40 & 51637 & 0.195 \\
21.06 & E & 560.79 & 06:55:25 & 29.71 & 51681 & 0.196 \\
22.06 & E & 561.79 & 06:55:57 & 30.01 & 43092 & 0.198 \\
23.06 & E & 562.75 & 05:57:46 & 30.30 & 50764 & 0.191 \\
24.06 & E & 563.75 & 06:00:11 & 30.60 & 50815 & 0.196 \\
22.06 & N   &  562.385  &   21:11:18 & 30.19 &  33442  & 0.188 \\
22.06 & N    &  562.486  &   23:36:57 & 30.22 &  27658  & 0.189 \\
27.06 & N    &  567.381  &   21:06:22 & 31.70 &  47415  & 0.193 \\
27.06 & N    &  567.484  &   23:34:15 & 31.73 &  33571  & 0.195 \\
28.06 & N   &  568.398  &   21:31:07 & 32.01 &  44240 &  0.192 \\
\\
\multicolumn{7}{c}{Map 4 (2016.54)} \\
02.07 & N    &  572.404  &   21:40:15 & 33.22 &  43145  &	0.191  \\
03.07 & N    &  573.399  &   21:32:19 & 33.52 &  54490  &	0.190 \\
05.07 & N    &  575.447  &   22:41:46 & 34.14 &  43655  &	0.189 \\
06.07 & N    &  576.385  &   21:12:21 & 34.42 &  26389  &	0.189 \\
08.07 & N    &  578.394  &   21:26:08 & 35.03 &  53673  &	0.191 \\
09.07 & N    &  579.427  &   22:13:34 & 35.34 &  45543  &	0.190 \\
14.07 & N    &  584.449  &   22:45:55 & 36.85 &  36385  &	0.190 \\
15.07 & N    &  585.440  &   22:32:28 & 37.15 &  42005  &	0.189 \\
19.07 & N    &  589.372  &   20:54:58 & 38.34 &  43063  &	0.188 \\
25.07 & N    &  595.366  &   20:47:27 & 40.15 &  49559  &	0.187 \\
26.07 & N    &  596.385  &   21:15:28 & 40.46 &  42596  &	0.187 \\
\hline
\hline
\end{tabular}
\end{table}

\label{lastpage}
\end{document}